\documentclass{ws-procs975x65}

\def\beq{\begin{equation}}
\def\eeq{\end{equation}}

\begin{document}

\title{The Vertex Expansion in the Consistent Histories Formulation of\\ Spin Foam Loop Quantum Cosmology}
\author{David Craig$^*$, Parampreet Singh$^\dagger$}

\address{$^*$Department of Chemistry and Physics, Le Moyne College\\
Syracuse, New York 13214, USA\\ \vskip0.2cm
$^\dagger$ Department of Physics, Louisiana State University\\
Baton Rouge, Louisiana 70810, USA}

\begin{abstract}
Assignment of consistent quantum probabilities to events in a quantum universe
is a fundamental challenge which every quantum cosmology/gravity framework
must overcome.
In loop quantum cosmology, this issue leads to a fundamental question: What is
the probability that the universe undergoes a non-singular bounce?  Using the
consistent histories formulation, this question was successfully answered
recently by the authors for a spatially flat FRW model in the canonical
approach.  In this manuscript, we obtain a covariant generalization of this
result.  Our analysis is based on expressing loop quantum cosmology in the
spin foam paradigm and using histories defined via volume transitions to
compute the amplitudes of transitions obtained using a vertex expansion.  We
show that the probability for bounce turns out to be unity.
\end{abstract}


\bodymatter


\section{Introduction}
Obtaining a consistent notion of probabilities is a fundamental problem in
quantum cosmology.  The usual Copenhagen interpretation is inadequate for a
quantum universe since there is no possible division of a quantum cosmos into
an observed quantum system and a classical external observer.  Measurement by
the latter or a classical detector which plays an essential role in Copenhagen
interpretation by killing the interference between alternative histories and
hence leads to the assignment of probabilities, no longer remains meaningful
in a closed quantum system.  The consistent histories approach is based on
generalizing the above procedure of eliminating the interference between
alternative histories without the baggage of observer based measurement.
\cite{griffiths,omnes,hartle}  Instead, this task is achieved by a
decoherence functional which measures the interference between different
alternatives.  These sets of alternatives define histories, specified by class
operators constructed from a time sequence of projection operators.  In case
the interference vanishes, histories decohere and consistent probabilities can
be assigned.

Recently, we have used the consistent histories formulation to compute the
probability for a non-singular bounce to occur in loop quantum cosmology
(LQC).  \cite{consistent-lqc,consistent-drev} Loop quantization of various
cosmological spacetimes has been performed,\cite{as} and the result that these
cosmological models bounce at small volume, first seen for the spatially flat
isotropic model, \cite{aps} has been generalized in various directions.
However, in most of the works, analysis of the quantum theory essentially
stops with computation of expectation values of observables and associated
fluctuations.  The consistent histories approach has filled an important gap
in LQC by providing precise answers to compute quantum probabilities in LQC.
The probability of a bounce in the spatially flat isotropic model sourced with
a massless scalar turns out to be unity, \cite{consistent-lqc} in striking
contrast to the Wheeler-DeWitt theory where the probability of a bounce is
zero for an arbitrary superposition of expanding and contracting universes
(analogous to Schr\"{o}dinger's cat states).  \cite{consistent-wdw} Instead,
the probability of a Wheeler-DeWitt universe to encounter a singularity turns
out to be unity.  So far, these computations were performed in the
conventional canonical formulation of LQC using an exactly solvable model.
\cite{acs} The goal of this work is to extend these results to a covariant
formulation using a spin foam avatar of LQC. \cite{consistent-sf} Our analysis
is based on earlier works building on this relationship between LQC and spin
foams.  \cite{ach1,ach2,ach3,cgo}

The manuscript is organized as follows.  In Sec.\ II, we summarize the primary
construction expressing LQC in langauge of spin foams.
\cite{ach1,ach2,ach3,cgo}  Starting from the quantum Hamiltonian constraint
$\hat C$ in solvable LQC, \cite{acs} we write the Hadamard propagator treating
$\hat C$ as a Hamiltonian.  The Hadamard propagator is equal to the physical
inner product obtained using group averaging, and can be written as a sum over
volume transitions by a careful rearrangement of summation in a vertex
expansion of the inner product.  In Sec.\ III, computation of the theory's
class operators and probabilities is summarized (see Craig and Singh
(2016)\cite{consistent-sf} for details).  Using the Hadamard propagator,
decoherence functionals are computed and the probability of a bounce is shown
to be unity.  The probability of a singularity to occur is found to be zero.
We conclude with a brief summary in Sec.\ IV.

\section{Sum over histories in exactly solvable LQC}
The spatially flat isotropic and homogeneous model in LQC offers a possibility
for an exactly solvable quantum theory.  If the lapse is chosen to be the
physical volume of the universe, the quantum Hamiltonian constraint simplifies
for the case of the massless scalar field $\phi$, and one obtains solvable LQC
\cite{acs} (sLQC).  In the volume representation, the action of the quantum
Hamiltonian constraint is given by\cite{acs}
\begin{equation}
\hat{C} \Psi(\nu,\phi) = - \left(\partial_{\phi}^2+\hat\Theta \right) \Psi(\nu,\phi) = 0 ~,
\end{equation}
where $\Theta$ is a positive definite and essentially self-adjoint operator with the following action:
\begin{eqnarray}
\Theta\,\Psi(\nu,\phi) &=& \nonumber  -\frac{3\pi G}{4\lambda^2} \bigg[ 
\sqrt{|\nu(\nu+4\lambda)|}|\nu+2\lambda|\Psi(\nu+4\lambda,\phi) - 2\nu^2\Psi(\nu,\phi) \\ && \hskip-1.5cm\nonumber 
+ \sqrt{|\nu(\nu-4\lambda)|}|\nu-2\lambda|\Psi(\nu-4\lambda,\phi) \bigg]  ~. ~~~ 
\end{eqnarray}
Here $\nu$ labels the eigenvalues of the volume operator: $\hat V |\nu\rangle
= 2 \pi \gamma |\nu| \mathrm{\ell_P}^2 |\nu\rangle$, and $\lambda = (4 \pi
\sqrt{3} \gamma^2 \mathrm{\ell_P}^2)^{1/2}$ with $\gamma$ as the
Barbero-Immirzi parameter.  The resulting quantum theory mimics the one for
the Klein-Gordon theory, and as in the latter, we have a superselection of the
positive and negative frequency states.  The physical Hilbert space can be
chosen as the one for the positive frequency states $\Psi^{(+)}$, which
satisfy
$- i \partial_\phi \Psi^{+}(\nu,\phi) = \sqrt{\Theta} \Psi^{+}(\nu,\phi) $.
For the positive frequency states, the transition amplitude between
$|\nu_i,\phi\rangle$ and $|\nu_f,\phi\rangle$ is given by the physical inner
product.  To cast LQC in the picture of sum over histories, it is useful to
work directly with these kinematical states, $|\nu_i,\phi_i\rangle$, which are
analogs of the spin networks in the spin foam framework.  The inner product
between two such spin network states, $|\nu_i, \phi_i\rangle$ and
$|\nu_f,\phi_f\rangle$, obtained using group averaging procedure, is given by
an analog of the Hadamard propagator $G_H$ \cite{cgo}
\begin{equation}
G_{\mathrm{H}}(\nu_f,\phi_f;\nu_i,\phi_i) =  \int_{-\infty}^{\infty} d\alpha\, \langle\nu_f,\phi_f|{e^{i\alpha\hat{C}}}|\nu_i,\phi_i\rangle ~.
\end{equation}
The integrand of the above integral is identified as the amplitude of transition $A(\nu_f,\phi_f;\nu_i,\phi_i;\alpha)$ which due to the separable form the 
Hamiltonian constraint can be written as 
\begin{equation}
A(\nu_f,\phi_f;\nu_i, \phi_i; \alpha) = A_\phi(\Delta \phi; \alpha) A_{\Theta}(\nu_f,\nu_i;\alpha) ~.
\end{equation}
Here
\begin{equation}
A_{\phi}(\Delta\phi;\alpha) = \langle \phi_f| {e^{i\alpha p_{\phi}^2}} |\phi_i\rangle, ~~\mathrm{and} ~~A_{\Theta}(\nu_f,\nu_i;\alpha) = \langle \nu_f| {e^{-i\alpha\Theta}} |\nu_i\rangle ~.
\end{equation}
The amplitude $A_\phi$ can be found easily using eigenfunctions of $\hat p_\phi$:
\begin{equation}
 A_\phi(\Delta \phi; \alpha) = \int d p_\phi \, {e^{i\alpha p_{\phi}^2}} {e^{i\Delta \phi p_\phi}} ~.
\end{equation}
On the other hand, finding the corresponding $A_{\Theta}(\nu_f,\nu_i;\alpha)$
is far more non-trivial.  Following Feynman's procedure, one considers a
division of the ``time'' interval $\Delta\phi$ in $N$ parts of length
$\epsilon/N$, with each sequence of intermediate volumes
corresponding to a ``history'' 
$(\nu_f,...,\nu_i)$.  Summing over these histories provides a transition
amplitude
\begin{equation}
 A_\Theta(\nu_f,\nu_i;\alpha) = \sum_{\bar\nu_{N-1},...\bar\nu_1} \langle\nu_f|e^{-i \epsilon \Theta}|\bar\nu_{N_1}\rangle .... \langle\bar\nu_1|e^{-i \epsilon \Theta}|\nu_i \rangle
\end{equation}
where we have used the resolution of the identity at each time step. It turns out that the limit $N \rightarrow \infty$ is tricky in the polymer representation of geometry which yields $\langle \nu_i | \nu_j \rangle = \delta_{ij}$ in contrast to a Dirac delta in the Fock representation. The result of this difference is that a na\"{i}ve sum to continuum limit is equivalent to considering $\epsilon^N$ with $\epsilon$ vanishing, leading to a null amplitude.
 
To take the continuum limit, the sum is rearranged in the spirit of the spin
foam vertex expansion.  The key idea is to group histories in terms of the
number of volume transitions, where each history satisfies the condition that
in a transition, a volume eigenvalue can not be repeated immediately.
However, a volume eigenvalue can return after a distinct volume transition.
For $m$ such transitions allowed in the fixed value of the time interval
$\Delta\phi$, the gravitational amplitude can be written as a reorganized sum
over the number of volume transitions $m$ of a sum over allowed paths with
exactly $m$ intermediate transitions, \cite{ach1,ach2}
\begin{equation}\label{ATheta}
A_\Theta(\nu_f,\nu_i;\alpha) = \lim_{N\rightarrow \infty} \sum_{m=0}^N \sum_{{\nu_{m-1},\ldots,\nu_1}\atop{\nu_m\neq \nu_{m-1}}} A(\nu_f,\nu_{m-1},...\nu_1,\nu_i;\alpha) ~.
\end{equation}
 
The Hadamard propagator 
\begin{equation} \label{GH_amp}
G_{\mathrm{H}}(\nu_f,\phi_f;\nu_i,\phi_i) =   \int_{-\infty}^{\infty} d\alpha\, 
A_{\mathrm{\phi}}(\Delta\phi;\alpha) A_{\Theta}(\nu_f,\nu_i;\alpha)
\end{equation}
can be computed by interchanging the integral with the summation in
eq.(\ref{ATheta}) by analogy of the procedure for a different propagator.
\cite{ach2}

A useful way to write the Hadamard propagator is to express it as a sum of positive and negative frequency components,
\begin{eqnarray}
G_{\mathrm{H}}(\nu_f,\phi_f;\nu_i,\phi_i) &=& \nonumber  \int_{-\infty}^{\infty}\frac{dk}{2\omega_k}\,  
[e^{+i\omega_k\Delta\phi} + e^{-i\omega_k\Delta\phi}]\,
e^{}_{k}(\nu_f) e^{}_{k}(\nu_i)^* \\
&=& \nonumber G^+_{\mathrm{H}}(\nu_f,\phi_f;\nu_i,\phi_i) + G^-_{\mathrm{H}}(\nu_f,\phi_f;\nu_i,\phi_i) ~.
\end{eqnarray}
Here $e_k$ are the eigenfunctions of the $\Theta$ operator with eigenvalues $\omega_k^2$.
Using the vertex expansion in eq.(\ref{GH_amp}), we obtain 
\begin{eqnarray}
 G_{\mathrm{H}}(\nu_f,\phi_f;\nu_i,\phi_i) &=& \nonumber \sum_{m=0}^{\infty} \sum_{{\nu_{m-1},\ldots,\nu_1}\atop{\nu_m\neq \nu_{m-1}}} \bigg[
A^{+}_m(\nu_{f},\nu_{m-1},\ldots,\nu_{1},\nu_{i};\Delta\phi) \\ && ~~~~~~~~~~~~~~~~~~~~ + A^{-}_m(\nu_{f},\nu_{m-1},\ldots,\nu_{1},\nu_{i};\Delta\phi) \bigg]
\end{eqnarray}
where the path amplitude  associated with the path 
$(\nu_{f},\nu_{m-1},\ldots,\nu_{1},\nu_{i})$ 
with $m$ transitions, where $\nu_{0}=\nu_i$ and $\nu_{m}=\nu_f$, such that there are $p$ unique volumes $(w_{p-1},w_{p-2},\ldots,w_{1},\nu_{0})$ 
%
is given by\cite{consistent-sf}
\begin{eqnarray}
  A^{\pm}_m(\nu_{f},\nu_{m-1},\ldots,\nu_{1},\nu_{i};\Delta\phi)  &=& \nonumber 
\Theta_{\nu_f\nu_{m-1}}\ldots\Theta_{\nu_2\nu_1}\Theta_{\nu_1 \nu_i}  \\
&& \nonumber \times\prod_{k=1}^{p}\frac{1}{(d_k-1)!} \left(\frac{\partial}{\partial \Theta_{w_k w_k}}\right)^{d_k-1} 
\sum_{i=1}^{p} \frac{(2\sqrt{\Theta_{w_i w_i}})^{-1} e^{\pm i\sqrt{\Theta_{w_i w_i}}\Delta\phi}}{\prod_{{j=0}\atop{j\neq i}}^{p-1} (\Theta_{w_i w_i}-\Theta_{w_j w_j})} ~.
\end{eqnarray}
Note that $p \leq m+1$. The degeneracy of each volume $w_k$ in the given path is denoted by $d_k$,
hence $\sum_{k=0}^{p-1}d_k=m+1$. The amplitude over the sum of gauge histories obtained by treating $\hat C$ as the Hamiltonian can thus be obtained 
from the matrix elements $\Theta_{\nu_i \nu_j} = \langle\nu_j | \Theta | \nu_i \rangle$.

\section{Class operators and probabilities}
For a given family of volume histories $\{h\}$ associated with $m$ volume transitions for states with positive frequencies, the class operator is given by \cite{consistent-sf}
\begin{equation}
C_{h}(\nu_f,\phi_f;\nu_i,\phi_i) =
\sum_{m=0}^{\infty} \sum_{\{\nu_{k}\}\in h}
A^{\pm}_m(\nu_{f},\nu_{m-1},\ldots,\nu_{1},\nu_{i};\Delta\phi) ~.
\end{equation}
Summing over all the families of histories, we obtain the Hadamard propagator for positive frequency states,
\begin{equation}
\sum_{h} C_{h}(\nu_f,\phi_f;\nu_i,\phi_i) =
G^+_{\mathrm{H}}(\nu_f,\phi_f;\nu_i,\phi_i) ~.
\end{equation}

Using the properties of the Hadamard propagator, we are now equipped to answer questions about the probability of bounce (or lack of it) in this exactly solvable model of LQC. We start with picking a reference volume, labeled by $\nu^*$, and classify paths according to
whether $\nu_{i,f}$ is greater or less than $\nu^*$ at $\phi_{i,f}$. That is, 
whether $\nu_i\in\Delta\nu^*$ or $\nu_i\in\overline{\Delta\nu^*}$ (and
similarly for $\nu_f$).

The class operator for histories for which 
$\nu_f\in\Delta\nu_1$ and $\nu_i\in\Delta\nu_2$ is:
\begin{equation}
C_{\smash{\Delta\nu_1;\Delta\nu_2}}(\nu_f,\phi_f;\nu_i,\phi_i)
= 
G^+_{\mathrm{H}}(\nu_f,\phi_f;\nu_i,\phi_i) 
\delta_{\nu_f,\Delta\nu_1}  \delta_{\nu_i,\Delta\nu_2} ~.
\end{equation}
Thus, the class operator for a history which bounces is,
\begin{equation}
C_{\mathrm{bounce}}(\nu_f,\phi_f;\nu_i,\phi_i) \,=\,
C_{\overline{\Delta\nu^*};\overline{\Delta\nu^*}}. 
\end{equation}
On the other hand, the class operator for the alternative history that
the universe is found at small volume at either or both of $\phi_i$, $\phi_f$
is
\begin{eqnarray*}
C_{\mathrm{sing}}(\nu_f,\phi_f;\nu_i,\phi_i)  & = & 
G^+_{\mathrm{H}}(\nu_f,\phi_f;\nu_i,\phi_i)  
- C_{\mathrm{bounce}}(\nu_f,\phi_f;\nu_i,\phi_i)  \\
& = &
C_{\smash{\Delta\nu^*;\Delta\nu^*}} +
C_{\smash{\Delta\nu^*;\overline{\Delta\nu^*}}} +
C_{\smash{\overline{\Delta\nu^*};\Delta\nu^*} }.
\end{eqnarray*}

It is important to note that 
for \emph{any} fixed volume $\nu_f$, by the Riemann-Lebesgue lemma, since at
fixed volume the factors in the integrand multiplying
$\exp(i\omega_k\Delta\phi)$ are integrable functions of $k$, the propagators
vanish in the limit $\Delta\phi\rightarrow \infty$.  
As a result, 
all 
the class operators appearing in $C_{\mathrm{sing}}$ are zero in the limits
$\phi_i\rightarrow -\infty$ and $\phi_f\rightarrow +\infty$ for any finite 
$\nu^*$. Thus,
\begin{eqnarray*}
C_{\mathrm{sing}} &=& 
\lim_{\substack{\phi_i\rightarrow -\infty\\ \phi_f\rightarrow +\infty }}
C_{\mathrm{sing}}(\nu_f,\phi_f;\nu_i,\phi_i) = 0.
\end{eqnarray*}
In contrast, the class operator corresponding to a bounce 
\begin{equation}
C_{\mathrm{bounce}} =
\lim_{\substack{\phi_i\rightarrow -\infty\\ \phi_f\rightarrow +\infty }}
C_{\mathrm{bounce}}(\nu_f,\phi_f;\nu_i,\phi_i) =
\lim_{\substack{\phi_i\rightarrow -\infty\\ \phi_f\rightarrow +\infty }}
G_{\mathrm{H}}^+(\nu_f,\phi_f;\nu_i,\phi_i) 
\end{equation}
is not zero. Therefore, \emph{all} states are driven to infinite volume as 
$|\phi|\rightarrow\pm\infty$.

The probability for the bounce to occur, $p_{\mathrm{bounce}}$, is captured 
by the decoherence functional element $d(\mathrm{bounce},\mathrm{bounce})$:
\begin{equation}
p_{\mathrm{bounce}} = d(\mathrm{bounce},\mathrm{bounce}) = \langle\Psi|C^\dagger_{\mathrm{bounce}} C_{\mathrm{bounce}} |\Psi\rangle ~.
\end{equation}
Computation of the decoherence functional reveals that this probability is
unity, \cite{consistent-sf} whereas the elements
$d(\mathrm{sing},\mathrm{bounce})$ and $d(\mathrm{sing},\mathrm{sing})$
vanish.  \cite{consistent-sf} The probability that in any volume transition, a
zero volume eigenvalue is reached is zero.  In this sense, the probability for
a singularity to occur turns out to be vanishing.

\section{Summary} 
Expressing LQC in the language of spin foams \cite{ach1,ach2,ach3,cgo} opens a
new window to explore answers to questions about consistent histories in a
covariant framework.  Given the rigorous detailed construction of the quantum
theory available in sLQC, \cite{acs,craig} the underlying task to define class
operators and calculate the decoherence functional becomes tractable.  The
decoherence functional can be computed to find the histories which decohere
and assign respective probabilities to such histories.  This task has been
accomplished for the first time in covariant formulation of sLQC.
\cite{consistent-sf}  Using histories classified via number of volume
transitions, we find the probability for a bounce to be unity.  This confirms
with our previous results on consistent histories in LQC using the canonical
approach. \cite{consistent-lqc}  Generalization of these results to a
covariant avenue provides an opportunity to explore further connections
between the these issues in LQG/spin foams and conventional path integral
approaches, \cite{craig-hartle} and also promises to give insights on some
fundamental issues in spin foams.

\section*{Acknowledgments}
PS is supported by NSF grants PHY-1403943, PHY-1454832 and PHY-1503417.

\end{document}